\documentclass[lettersize,journal]{IEEEtran}
\usepackage[flushleft]{threeparttable}
\usepackage{amsmath,amsfonts}

\usepackage{cite}
\usepackage{array}
\usepackage[caption=false,font=normalsize,labelfont=sf,textfont=sf]{subfig}

\usepackage{etoolbox}

\usepackage[]{footmisc}
\usepackage{textcomp}
\usepackage{enumitem}
\usepackage{stfloats}
\usepackage{url}
\usepackage{verbatim}
\usepackage{graphicx}
\usepackage{multirow}
\usepackage{mathtools}
\usepackage{breqn}
\usepackage[flushleft]{threeparttable}
\usepackage{multicol}
\usepackage{array,colortbl,xcolor}
\usepackage{color,soul}
\usepackage{xcolor}
\usepackage{tikz}
\newcommand*\circled[1]{\tikz[baseline=(char.base)]{
            \node[shape=circle,draw,inner sep=0.4pt] (char) {#1};}}
\usepackage{mathtools}
\usepackage{amsmath,amsfonts}
\usepackage{amssymb,mathtools}
\usepackage{dashbox}%
\usepackage{algorithm}

\begin{document}
\title{COMET: Co-Optimization of {CNN Models} using Efficient-Hardware OBC Techniques}
\author{Boyang~Chen,  Mohd~Tasleem~Khan,~\IEEEmembership{Member,~IEEE},  George~Goussetis,~\IEEEmembership{IEEE Fellow},\\ 
Mathini~Sellathurai,~\IEEEmembership{IEEE Fellow}, Yuan~Ding,~\IEEEmembership{Member,~IEEE}, João~F. C.~Mota,~\IEEEmembership{Member,~IEEE},\\ and Jongeun Lee,~\IEEEmembership{Senior Member,~IEEE}}

\IEEEtitleabstractindextext{%
\begin{abstract}
Convolutional Neural Networks (CNNs) achieve remarkable accuracy in vision tasks, yet their computational complexity challenges low-power edge deployment. In this work, we present COMET, a framework of {CNN models} that employ efficient hardware offset-binary coding (OBC) techniques to enable co-optimization of performance and resource utilization. The approach formulates CNN inference using OBC representations applied separately to inputs (Scheme A) and weights (Scheme B), enabling exploitation of bit-width asymmetry. The shift–accumulate operation is modified by incorporating offset-term with the pre-scaled bias. Leveraging symmetries in Schemes A and B, we introduce four look-up table (LUT) techniques—parallel, shared, split, and hybrid—and evaluate their efficiency. Building on this foundation, we develop a general matrix multiplication core using the \textit{im2col} transformation for efficient CNN acceleration. {We consider LeNet-5 and All-CNN-C to demonstrate that the OBC-GEMM core efficiently supports modern workloads. Evaluation shows that COMET enables efficient FPGA deployment compared to state-of-the-art designs, with negligible accuracy loss, demonstrating its efficiency and scalability across diverse network architectures.}

\end{abstract}
\begin{IEEEkeywords}
Convolutional Neural Network (CNN), Field-Programmable Gate Array (FPGA), General Matrix-Multiply (GEMM), Look-up Table (LUT), Offset-Binary Coding (OBC).  
\end{IEEEkeywords}}

\maketitle
\IEEEdisplaynontitleabstractindextext

\IEEEpeerreviewmaketitle
\section{Introduction}
\IEEEPARstart{C}{onvolutional} neural networks (CNNs) learn features directly from images and outperform traditional methods in classification, detection, and segmentation tasks \cite{zhao2024review}. Due to their strong representational capability, they are widely used in applications such as industrial defect inspection and medical imaging \cite{khanam2024comprehensive}. {Modern CNNs incorporate deep topologies, residual and dense connections, attention mechanisms, normalization layers, depthwise separable convolutions, and multi-branch structures to improve performance} \cite{khan2020survey}. {However, this architectural complexity increases computational cost, as convolutions dominate the arithmetic workload in widely used architectures such as VGG-style}  \cite{wang2021avnc}, {ResNet} \cite{ren2021spatio}, {MobileNet} \cite{pan2020new}, {and attention-augmented models} \cite{bello2019attention}.

{Each convolution layer consists of numerous multiply–accumulate (MAC) operations, often reformulated using the \textit{im2col} transformation into matrix–matrix or matrix–vector multiplications in the form inner product computations (IPCs)} \cite{fornt2023energy}. {As network depth, channel count, and kernel size increase, these operations impose significant computational and memory demands, complicating deployment on low-power field programmable gate array (FPGA)-based platforms} \cite{zhao2025high, khan2026next}. Although FPGAs provide parallelism and reconfigurability, large MAC arrays quickly consume DSP resources, increase routing congestion, and raise power consumption, limiting scalability.

To improve resource utilization, many CNN accelerators adopt \textit{im2col}-based general matrix multiplication (GEMM) architectures to reuse shared MAC arrays~\cite{fornt2023energy,guo2024accelerating}. While this approach improves throughput, it also increases DSP and memory usage. Small CNNs can be deployed relatively easily on modest hardware, whereas larger CNNs typically require bigger FPGA devices to accommodate the increased compute and storage demands. For resource-constrained devices, replacing DSPs with LUTs and BRAMs offers a promising direction, although the trade-off between LUT resource pressure and BRAM utilisation becomes increasingly important as network complexity grows. Even advanced dataflow strategies such as~\cite{kim2023agamotto}, which maximise data reuse, remain fundamentally dependent on MAC units, thereby constraining overall efficiency and scalability.


Distributed arithmetic (DA) is a multiplierless approach suited for low-power, area-constrained hardware \cite{white2002applications, khan2022two}, replacing MAC operations in IPC with a pre-computed look-up table (LUT) and a shift-accumulate (SA) unit. This is achieved by representing the elements of a vector in an IPC either in two's complement (TC) \cite{khan2022high} or offset-binary coding (OBC) \cite{khan2018optimal}, with OBC halving the LUT size through symmetry. Recent DA-based LSTM optimizations use hardware LUTs \cite{yalamarthy2019low, khan2022architectural, alhartomi2023low, khan2024digit}. Parallel OBC-LUTs reduce critical path delay (CPD) but increase complexity \cite{yalamarthy2019low}, while serial LUTs with high-radix OBC/TC reduce memory but add latency \cite{khan2022architectural, khan2024digit}. Few DA-specific CNN or 1-point convolution implementations exist \cite{panwar2017modified, chen2019area, chen2023performance}, e.g., input pairing to shrink LUTs \cite{panwar2017modified} or in-memory optimization to reduce memory \cite{chen2019area}. For efficient CNN implementation on FPGA platforms, asymmetric quantization is effective: reducing the weight bit-width while keeping input precision fixed lowers power consumption at a modest accuracy cost \cite{krishnamoorthi2018quantizing}. However, this strategy has not yet been investigated in DA-based CNN architectures, particularly with respect to validation on FPGA platforms. Moreover, its impact on model accuracy and hardware efficiency across various CNN architectures has not been clearly established.


\begin{figure*}[t]
	\centering  \includegraphics[width=0.89\linewidth]{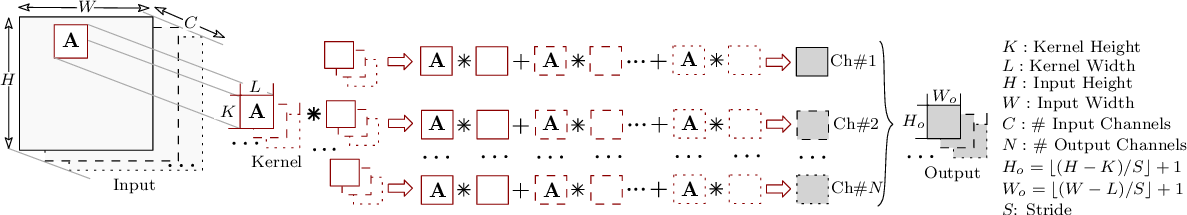}
	\caption{Illustration of the {\textit{im2col}} transformation of a 2D convolution into a GEMM operation, assuming no padding in the input feature map and stride=1.} \label{fig1}
\end{figure*}

{To address these challenges, we propose COMET, our co-optimization framework for CNN hardware acceleration. For evaluation purposes, we adopt LeNet-5} \cite{lecun1998gradient} and All-CNN-C \cite{springenberg2014striving} {as complementary benchmarks. LeNet-5, an early CNN with a compact structure and low parameter count, provides a controlled platform to analyze resource, performance, and accuracy trade-offs; it often requires lightweight modification for efficient deployment} \cite{imen2022fast}. {In contrast, All-CNN-C is a complex, modern, fully convolutional architecture, with strided convolutions replacing fully connected and pooling layers, stressing scalable IPCs without requiring modifications. Both networks rely on structured IPCs: dense layers use full weight matrices, and convolutional layers use sparse, weight-shared kernels. Through {\textit{im2col}}, convolutions are reformulated as GEMM-based IPCs. COMET reduces DSP usage and power consumption without altering network functionality, employing a streaming ({\textit{im2col}} + tiled) strategy and dynamic parallel LUTs to compute partial sums at runtime. Standard CNN kernels are executed as OBC-based GEMM operations within a scalable hardware architecture. Evaluation on LeNet-5 and All-CNN-C demonstrates that COMET generalizes across lightweight and convolution-dominated networks, validating both feasibility and scalability toward more complex CNNs. Our main contributions are as follows:}

\begin{itemize}
	\item We propose a \textit{im2col}-based formulation of the 2D CNN convolution layer using OBC representations of inputs (Scheme~A) and weights (Scheme~B), as initial part of the COMET framework.  	
	\item We introduce four LUT architectures based on efficient hardware OBC techniques and optimize the SA unit by unifying pre-scaled bias and offset terms, improving hardware efficiency.
	\item We explore and evaluate the design space of the OBC-GEMM framework using Scheme A and Scheme B based on the proposed LUT architectures, analyzing key design trade-offs.  
    \item We investigate the impact of fixed-point quantization on the accuracy of LeNet-5 {and All-CNN-C} models and integrate the proposed OBC-GEMM core into their convolutional kernels for hardware evaluation.  
    \item We compare the {proposed architectures integrated with LeNet-5 and All-CNN-C against state-of-the-art hardware accelerators, highlighting the improvements in resource–energy trade-offs.} 
\end{itemize}

The remainder of this paper is organized as follows. Section~II introduces the OBC formulation of a standard CNN. Section~III presents the proposed LUT architectures and their co-optimization trade-offs. Section~IV covers the evaluation of the OBC-GEMM module. Section~V describes the CNN models consideration and accuracy results. Section~VI provides comparison with state-of-the-art CNN accelerators, and Section~VII concludes the paper.

\begin{figure*}[t]
  \centering
  \includegraphics[width=0.88\linewidth]{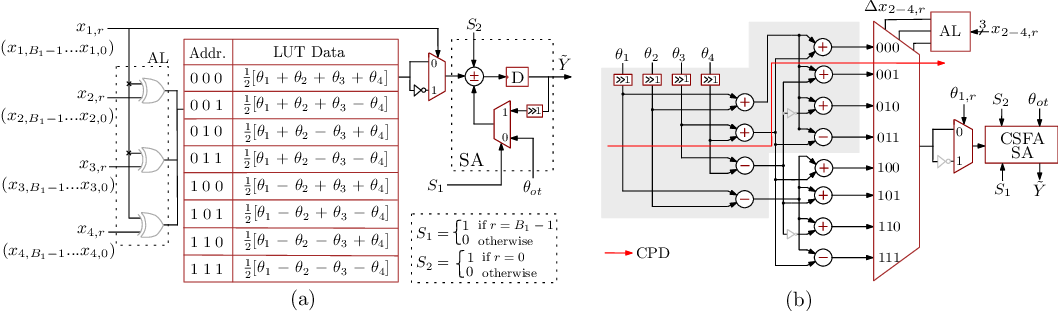}
  \caption{OBC-based IPC for $K=4$ with: (a) Traditional LUT and SA-unit \cite{khan2018optimal}, (b) Hardware LUT (w/o pipelining) and CSFA-based SA unit \cite{yalamarthy2019low}.}
  \label{fig2}
\end{figure*}
\section{CNN Formulation with OBC}
As shown in Fig.~\ref{fig1}, a standard CNN is a multi-layer pipeline where each layer transforms input feature maps of size $H \times W$ into higher-level representations using convolutional kernels of size $K \times L$ and biases. A layer with $C$ input and $N$ output channels computes each output by convolving all $C$ input channels with their respective kernels via a sliding-window operation (\textbf{A}) and summing across channels. This process is repeated for all $N$ outputs. For simplicity, a batch size of $1$ is assumed, though the formulation extends to multiple batches. Specifically, given an input tensor with $C$ channels, each with spatial dimensions $H\times W$, i.e., $X \in \mathbb{R}^{C\times H\times W}$, the output of a 2D convolution of $X$ with a filter tensor $\Theta \in \mathbb{R}^{N \times C \times K \times L}$ is a tensor $Y \in \mathbb{R}^{N\times (H+K) \times (W+L)}$, whose entries are 
\begin{equation}\label{eq1}
	Y_{n,h,w} = \sum_{c=1}^{C}\sum_{k=1}^{K}\sum_{l=1}^{L}
	X_{c,h+k,w+l}\Theta_{n,c,k,l} + \beta_{n},
\end{equation}
where $n=1, \ldots, N$, $h=1, \ldots, H+K$, $w=1,\ldots, W+L$, and $\beta_n \in \mathbb{R}$ represents a bias term. Note ~\eqref{eq1} describes the convolution computation for a single layer. Each product $X_{c,h+k,w+l} \Theta_{n,c,k,l}$ followed by its accumulation constitutes one MAC operation. Therefore, a single convolution layer requires roughly $N \times H \times W \times C \times K \times L$ MAC operations, and across multiple layers, total MACs can reach tens of millions, stressing compute and memory bandwidth~\cite{jiang2022high}. To mitigate this, we adopt the \textit{im2col} transformation~\cite{fornt2023energy}, which converts convolutions into tile-based GEMM, enabling the reuse of a single MAC array across layers. Applied to (\ref{eq1}), the transformation typically follows these steps:
\begin{itemize}
	\item For each output position $(h,w)$, a 3D patch of the input is extracted as
	\[
	\text{patch}_{h,w} = X[:,\,h:h+K,\,w:w+L] \in \mathbb{R}^{C\times K\times L}.
	\]
    \item Each extracted patch $\text{patch}_{h,w} \in \mathbb{R}^{C \times K \times L}$ is flattened into a column vector:
Each extracted patch $\text{patch}_{h,w} \in \mathbb{R}^{C \times K \times L}$ is flattened into a column vector
\[
\text{col}_X(h,w) \in \mathbb{R}^{C K L} \triangleq x_i,
\]
where $N_p \triangleq C K L$ denotes the patch dimension.

\item Each filter $\Theta_n \in \mathbb{R}^{C \times K \times L}$ is flattened into a row vector:
\[
\text{row}_\Theta(n) \in \mathbb{R}^{N_p} \triangleq \theta_n.
\]
	\item The convolution is expressed as an IPC of length $N_p=CKL$ between the flattened input patch and filter:  
	\begin{equation}\label{eq3}
		{Y} = \sum_{i=1}^{CKL} \theta_i \, x_i+\beta_i.
	\end{equation}
\end{itemize}

The resulting output is then reshaped into the proper order. For simplicity, we set $C=1$ and $L=1$ in the IPC formulation, though the analysis extends to any $C$ and $L$. Denoting the summation in (\ref{eq3}) as ${\tilde{Y}}$, we have:
\begin{equation}\label{eq4}
	{\tilde{Y}} = \sum_{i=1}^{K} \theta_i \, x_i.
\end{equation}
\noindent In (\ref{eq4}), the IPC in OBC form \cite{khan2022two} either the inputs $x_i$, a scheme that we refer as Scheme A, or the weights $\theta_i$, a scheme that we refer as Scheme B. For Scheme A, each input  \(x_i\) with bit-width of \(B_1\)-bits is first represented in TC form as 
\begin{equation}\label{eq:xi_tc}
x_i = -\,x_{i,0} + \sum_{r=1}^{B_1-1} x_{i,r}\,2^{-r}
\end{equation}
where $x_{i,r}\in\{0,1\}$ is the $r$th bit of $x_{i}$. In OBC representation, $x_{i}$ can be written as $x_{i}=\frac{1}{2}[x_{i}-(-x_{i})]$, which results (\ref{eq:xi_tc}) in 
\begin{equation}\label{eq:OBC_trick_x}
x_i=\frac{1}{2}\left[\sum_{r=0}^{B_1-1}\Delta x_{i,r}2^{-r}-2^{-(B_1-1)}\right]    
\end{equation}
with
\begin{equation}\label{eq6}
\Delta x_{i,r} = (-1)^{\lfloor r/(B_1-1) \rfloor}\left(x_{i,r}-\bar x_{i,r}\right)    
\end{equation}
where $\bar x_{i,r}$ is the ones complement of $x_{i,r}$ and $\lfloor \cdot \rfloor$ is the floor function. Substituting \eqref{eq:OBC_trick_x} and \eqref{eq6} into \eqref{eq4}, we have
\begin{equation}\label{eq:y_n_da}
{\tilde{Y}}
= \sum_{r=0}^{B_1-1}
      \Bigl(\sum_{i=1}^{K}\tfrac12\,\theta_i\,\Delta x_{i,r}\Bigr)\,2^{-r}
    - \Bigl(\tfrac12\sum_{i=1}^{K}\theta_i\Bigr)\,2^{-(B_1-1)}
\end{equation}
\noindent Define
\begin{equation}\label{eq:p_defs}
{\tilde{Y}}_{r}=\tfrac12\sum_{i=1}^{K}\,\theta_i\,\Delta x_{i,r} \hspace{0.2cm} \textnormal{and},  \hspace{0.2cm} \theta_{ot} = -\tfrac12 \sum_{i=1}^{K}\theta_i
\end{equation}
Substituting (\ref{eq:p_defs}) into (\ref{eq:y_n_da}), we  arrive at
\begin{equation}\label{eq:y_da_final}
{\tilde{Y}} = \sum_{r=0}^{B_1-1} {\tilde{Y}}_r\,2^{-r}
  + \theta_{ot}\,2^{-(B_1-1)}
\end{equation}
Substituting (\ref{eq:y_da_final}) into (\ref{eq4}) and using (\ref{eq3}), we get
\begin{equation}\label{eq10}
{Y}  = \sum_{r=0}^{B_1-1} {\tilde{Y}}_r\,2^{-r}
  + (\theta_{ot}+\beta_i 2^{B_1-1})2^{-(B_1-1)}
\end{equation}
where $\tilde\theta_{ot}=\theta_{ot}+\beta_i 2^{B_1-1}$ indicates that an offset term is combined with a scaled bias term. Thus, (\ref{eq10}) leads to 
\begin{equation}\label{eq11}
{Y}  = \sum_{r=0}^{B_1-1} {\tilde{Y}}_r\,2^{-r}
  + \tilde\theta_{ot}2^{-(B_1-1)}
\end{equation}

\noindent Each LUT entry represents a combination of weights with input bit-slices 
\((x_{n,B_1-1}, \ldots, x_{n,0}; 1 \le n \le K)\), ordered from least to most significant bit, with bit-slices used as addresses at runtime. Outputs are accumulated via a SA operation over $B_1$ clock cycles to compute $\tilde{Y}$, as shown in Fig.~\ref{fig2}(a). The offset term $\theta_{ot}$ is the twos complement of the LUT content at address 0, loaded at the start of shift-accumulation, as shown in Fig.~\ref{fig2}(a). Traditional LUTs become inefficient for large $K$ due to access time. Dynamic generation using adders and multiplexers~\cite{yalamarthy2019low} lacks scalability and adds hardware overhead: $2^{K}(1 - 2^{-K/2})$ adders/subtractors, $2^{K-1}-1$ 2-to-1 multiplexers, and CPD of $\log_2 K \cdot T_A + (K-1) \cdot T_{MX}$, where $T_A$ and $T_{MX}$ are adder and multiplexer delays. Note carry-save full adders (CSFAs) based SA unit has been employed to reduce this delay.

Following (\ref{eq11}), $Y$ can be computed using Scheme B, where each weight $\theta_i$ in OBC form with bit-width $B_2$ is processed as in (\ref{eq:xi_tc})–(\ref{eq11}), yielding:
\begin{equation}\label{eq12}
{Y}  = \sum_{r=0}^{B_2-1} {{\tilde{Z}}}_r\,2^{-r}
  + {\tilde{x}}_{ot}2^{-(B_2-1)}
\end{equation}
where the variables in (\ref{eq12}) are defined as follows:
\begin{align*}
{\tilde{Z}}_{r} &= \tfrac{1}{2}\sum_{i=1}^{K} \,x_i\,\Delta \theta_{i,r}\\
\Delta \theta_{i,r} &= (-1)^{\left\lfloor r/(B_2-1) \right\rfloor}\left(\theta_{i,r}-\bar{\theta}_{i,r}\right) \\
\tilde{x}_{{ot}} &= x_{{ot}} + \beta_i\,2^{B_2-1}, \hspace{0.2cm} \textnormal{and} \hspace{0.2cm} x_{{ot}} = -\tfrac{1}{2} \sum_{i=1}^{K} x_i.   
\end{align*}
\noindent Note that Scheme~B shares the same structure as Scheme~A, except that the LUT content and SA clock cycles may differ ($B_1 \neq B_2$). The schemes differ only in which component is represented in OBC form, leveraging the weight–input bit-width asymmetry to offer hardware implementation trade-offs.

\section{Proposed Architecture}
\subsection{System Overview}
\begin{figure}[t]
  \centering  \includegraphics[width=0.82\linewidth]{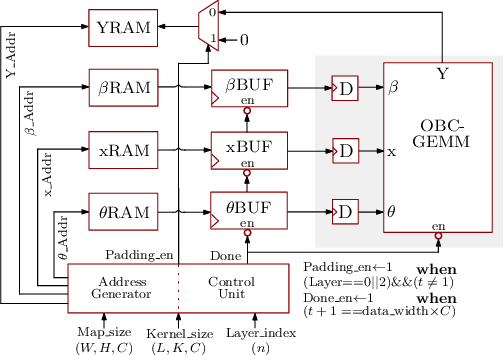}
  \caption{System-level diagram of the proposed CNN accelerator. Note $\theta$RAM, xRAM, $\beta$RAM, YRAM are on-chip memory blocks and $\theta$BUF, xBUF, $\beta$BUF are their associated buffers.}
  \label{fig3a}
\end{figure}

{The OBC-GEMM core, incorporating Scheme A and Scheme B, is first mapped onto hardware-optimized architectures. Subsequently, CNN models such as LeNet-5 and All-CNN-C are employed as representative case studies to evaluate computational efficiency across diverse CNN topologies.} An overview of the proposed CNN accelerator is shown in Fig.~\ref{fig3a}. The architecture includes the OBC-GEMM compute core, on-chip memory blocks ($\theta$RAM, xRAM, $\beta$RAM, YRAM) with their buffers ($\theta$BUF, xBUF, $\beta$BUF), an \textit{im2col} address generator, and a centralized control unit. A finite state machine (FSM) coordinates the control unit and address generator to manage dataflow, computation, and state transitions. The OBC-GEMM core, based on hardware-LUT architectures, is embedded into the accelerator.

At the start of each layer, a configuration word is loaded, encoding kernel dimensions ($C \times K \times L$), stride ($S$, with $S=1$ for normal and $S=2$ for downsampling), padding type/amount ($P$, with $P=0$ for no padding and $P=1$ for one-sided zero padding), input/output channels ($C$/$N$), and bit-width ($B_1$ or $B_2$). A cycle counter drives the \textit{im2col} address generator, computing read addresses dynamically and injecting zeros for padding. Inputs are fetched from RAM and streamed into the OBC-GEMM engine, while the next chunk is prefetched via ping–pong buffers. Partial or final sums are written back to RAM at the generated addresses, ensuring correct spatial alignment. This repeats for all layers, forming a fully streamed, stall-free pipeline.
 

\subsection{Address Generator and Control Unit}
\begin{figure}[t]
  \centering  \includegraphics[width=0.85 \linewidth]{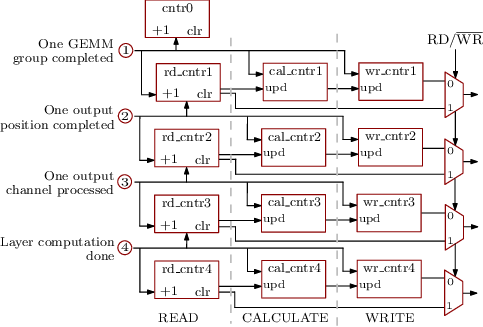}
  \caption{Hierarchical counter mechanism enabling address generation across read, calculate, and write stages.}
  \label{fig3b}
\end{figure}

At the core of the address generator is a set of nested counters managed by the FSM, handling address generation and read/write scheduling. As shown in Fig.~\ref{fig3b}, the architecture includes a dedicated \texttt{cntr0} and three hierarchical counter groups for read (\texttt{rd\_cntr\#$m$}), calculation (\texttt{cal\_cntr\#$m$}), and write (\texttt{wr\_cntr\#$m$}) operations. Each group has four levels of counters indexed by \( m \in \{1,4\} \), corresponding to tiling, spatial position \((h,w)\), output channel \((n)\), and layer index. \texttt{cntr0} tracks the OBC-GEMM clock cycle. Signals \(\circled{1}\)–\(\circled{4}\) in Fig.~\ref{fig3b} indicate the hierarchical carry mechanism, enabling stepwise progression and value transfer between read, calculation, and write stages.

In this hierarchical carry mechanism, rippling primarily affects the read counters. The clock cycle counter (\texttt{cntr0}) generates sequential memory addresses for one tile-level OBC-GEMM operation. When it reaches its upper bound, all data for the current tile are loaded, carry \(\circled{1}\) is asserted, \texttt{cntr0} resets, and \texttt{rd\_cntr1} (tile index) increments to start the next tile. When \texttt{rd\_cntr1} completes, carry \(\circled{2}\) resets it and increments \texttt{rd\_cntr2} (spatial position). Similarly, \(\circled{3}\) resets \texttt{rd\_cntr2} and increments \texttt{rd\_cntr3} (output channel), and \(\circled{4}\) resets \texttt{rd\_cntr3} and increments \texttt{rd\_cntr4} (layer index) to begin the next layer.

\begin{figure}[t]
  \centering\includegraphics[width=0.83\linewidth]{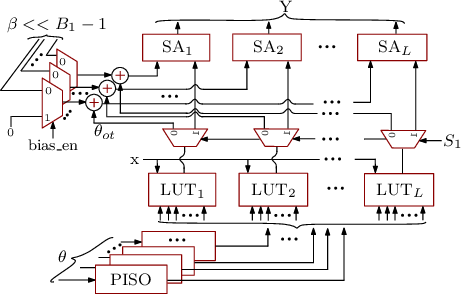}
  \caption{Top-level schematic of the proposed OBC-GEMM core.}
  \label{fig3c}
\end{figure}
Unlike read counters, the calculation and write-back counters do not ripple. Instead, when carry \(\circled{i}\) is asserted, \texttt{rd\_cntr\#m} values propagate to \texttt{cal\_cntr\#m} and \texttt{wr\_cntr\#m}, reflecting the dataflow: read \(\rightarrow\) compute \(\rightarrow\) write. This synchronization but decoupling enables lightweight task-level pipelining, allowing parallel read, compute, and write without interference, boosting throughput. Together, the counters define the complete computation context. For example, \(\texttt{cntr0}=11\), \(\texttt{rd\_cntr1}=2\), \(\texttt{rd\_cntr2}=5\), \(\texttt{rd\_cntr3}=1\), \(\texttt{rd\_cntr4}=0\) corresponds to fetching the 12\textsuperscript{th} data group in tile 3, spatial position 5, output channel 1, layer 0. A mapping function combines these counters to generate precise memory addresses. Read counters select memory positions, the address generator fetches tile data from \(\theta\)RAM, xRAM, and \(\beta\)RAM into buffers, calculation counters fix output positions, and write counters generate YRAM addresses. The architecture also supports bias addition—applied only at the last tile of a spatial location—and padding, inserted by monitoring \texttt{cntr0} to detect invalid regions at output boundaries.

\begin{figure*}[t]
  \centering
  \includegraphics[width=1\linewidth]{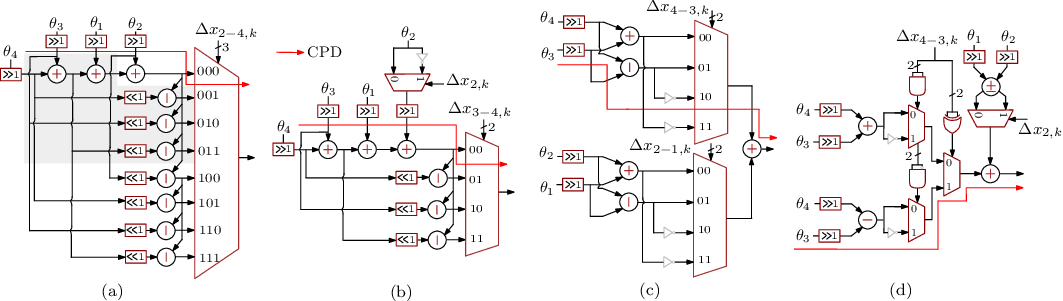}
  \caption{LUT architectures for $K=4$ based on the proposed techniques: (a) Parallel, (b) Shared, (c) Split, and (d) Hybrid.}
  \label{fig4}
\end{figure*}

\subsection{OBC-GEMM Core}
The OBC-GEMM core comprises $L$ LUTs of size $K$, $L$ SA units, and $L$ PISO units of size $K$ for handling input or weight bit-slices, as shown in Fig.~\ref{fig3c}. LUTs dynamically generate partial sums via adders and multiplexers, while SA units combine them efficiently. PISO units serialize inputs or weights to match the dataflow of Scheme A or B. A key optimization is a single adder for both offset and bias computation, eliminating separate bias logic and reducing hardware complexity, area, and power.

\subsubsection{Parallel LUT}
This initial architecture uses fewer adders than the one shown in Fig.~\ref{fig2}(b) and generates all LUT contents from the value at address $0$. Chain adders reuse the partial sum $\theta_3 + \theta_4$, which appears in the LUTs at addresses 3, 4, and 7, as illustrated in Fig.~\ref{fig4}(a). This sum is added to $\theta_1$, present in all addresses, and then added or subtracted from $\theta_2$ to produce all OBC combinations. The LUT at address 1 can be obtained by subtracting $2\theta_2$ (via a right shift) from address 0. This process continues with subtractors for all addresses. While CPD grows linearly with $K$, the number of adders and multiplexers grows exponentially. To limit this, $K$ can be factored as $K = p \times q$, restricting the exponential growth to $q$, with higher-order $K$ implemented using an adder tree of depth $\log_2 p$.

\begin{table}[t]
	\caption{{Theoretical Hardware and Time Complexities\\ of the Proposed LUT Architectures with $K=p\times q$}}\label{table1}
	\centering
	\resizebox{1\linewidth}{!}{%
		\begin{threeparttable}
			\begin{tabular}{l|c|c|c}
				\hline
				{{\bf{Technique}}} & {{\bf{Adders}}} & {{\bf{2-to-1 Muxes}}} & {{\bf{CPD}}}  \\ \hline
				Parallel & $(2^{q-1}+q-2)p+p-1$  & $(2^{q-1}-1)p$  & $(q+\textnormal{log}_2p)T_{A}$   \\ \hline
				Shared & $(2^{q-2}+q-2)p+p-1$  & $2^{q-2}p$  & $(q+\textnormal{log}_2p)T_{A}$   \\ \hline
				Split & $(2\cdot(2^{q/2-1}-1)+1)p+p-1$  & $2\cdot(2^{q/2})p$  & $(q/2+1+\textnormal{log}_2p)T_{A}+T_{MX}$   \\ \hline
				Hybrid & $qp+p-1$  & $3(q-2)+1$  & $(2+{\textnormal{log}}_2p)T_{A}+2T_{MX}$   \\ 
				\hline
			\end{tabular}
			\begin{tablenotes}
				\item Hybrid LUT has also $qp/2$--AND and $qp/4$--XOR gates to generate select signals for 2-to-1 multiplexers. $T_{A}$ and $T_{MX}$ are the computational delays of an adder and a 2-to-1 multiplexer respectively. 
			\end{tablenotes}
		\end{threeparttable}
	}
\end{table}

\subsubsection{Shared LUT}
The shared LUT approach builds on the parallel LUT by sharing generated contents to reduce partial sums, as shown in Fig.~\ref{fig4}(b). For example, LUT contents at addresses 4--7 can be derived by loading $-\theta_2$ instead of $\theta_2$ via a 2-to-1 multiplexer, lowering computational cost. Directly applying this to Fig.~\ref{fig2}(b) is challenging due to exponential adder growth. By sharing partial sums across mirrored addresses (e.g., 0--3 and 4--7), the shared LUT design reduces computations by nearly 50\%.

\subsubsection{Split LUT}
In both parallel and shared LUT techniques, adders and multiplexers grow exponentially. While shared LUTs halve computations, exponential scaling remains. To address this, the LUT is split into two parts.The first $P$ terms in the summation defined by $\tilde{Y}_r$ in (\ref{eq:p_defs}) are implemented using a parallel LUT with $(2^{P-1}-1)+(P-1)$ adders and a $2^{P-1}$-to-1 multiplexer (or $2^{P-1}-1$ 2-to-1 MUXes). The remaining $K-P$ terms are computed with $(2^{K-P-1}-1)+(K-P-1)$ adders and a $2^{K-P-1}$-to-1 multiplexer.
One adder combines the two partial results. Thus, total adders are $(2^{P-1}+2^{K-P-1}-2)+(K-2)$ and 2-to-1 MUXes are $2^{P-1}+2^{K-P-1}-2$. Decreasing $P$ reduces first-part complexity but increases the second, and vice versa; minimizing total hardware shows the optimum at $P=K/2$. The proposed LUT architecture based on this technique is shown in Fig.~\ref{fig4}(c). It can be observed that the chain of adders is replaced by a single adder for the first $K/2$ section, while the other term is computed using a subtractor. The remaining two partial sums are simply the twos complements of these results. A similar structure applies to the second $K/2$ section as well.

\subsubsection{Hybrid LUT}
Although the split LUT reduces computational challenges of parallel and shared LUTs, it still grows exponentially with $K$. To address this, a hybrid LUT is proposed shown in Fig.~\ref{fig4}(d), pairing consecutive weights and using minimal logic to generate partial sums. Here, hybrid refers to reducing large-scale parallelism via serial-like weight pairing. Adders and subtractors compute $\theta_{K-i+1} + \theta_{K-i}$ and $\theta_{K-i+1} - \theta_{K-i}$ for $i \in \{1, \dots, K-2\}$ and can be shared, while the last pair uses a conditional adder to minimize resources. Each weight pair requires only one adder or subtractor, yielding linear resource scaling. The selection lines for the 2-to-1 multiplexers to select between $\theta_{K-i+1} + \theta_{K-i}$ and $\theta_{K-i+1} - \theta_{K-i}$—can be derived from the observation that the contents at the $0^{\textnormal{th}}$ and $3^{\textnormal{rd}}$ address locations are two’s complements of each other, as are the $1^{\textnormal{st}}$ and $2^{\textnormal{nd}}$, and similarly for the second half, as illustrated in Fig.~\ref{fig2}(a).

\subsection{Analysis of Co-Optimization Trade-offs in LUTs}
\subsubsection{Hardware and Time Complexities}
Table~\ref{table1} lists theoretical hardware complexity (in terms of adders and 2-to-1 multiplexers) and time complexity (in terms of CPD) for the proposed LUT architectures, estimated using $K = p \times q$ for fair comparison. The Parallel LUT has the highest hardware cost due to exponential growth with $q$. The Shared LUT reduces adders by reusing partial sums but maintains similar CPD. The Split LUT further lowers adders at the cost of more multiplexers. The Hybrid LUT achieves the best balance, with linear adder scaling, reduced multiplexers, and low CPD due to shallow adder trees. For comparison, $q = 4$ ($K = 4p$), and all LUTs are implemented on FPGA with 8-bit symmetric operands, as detailed in the next subsection.

\begin{figure*}
    \centering    \includegraphics[width=0.92\linewidth]{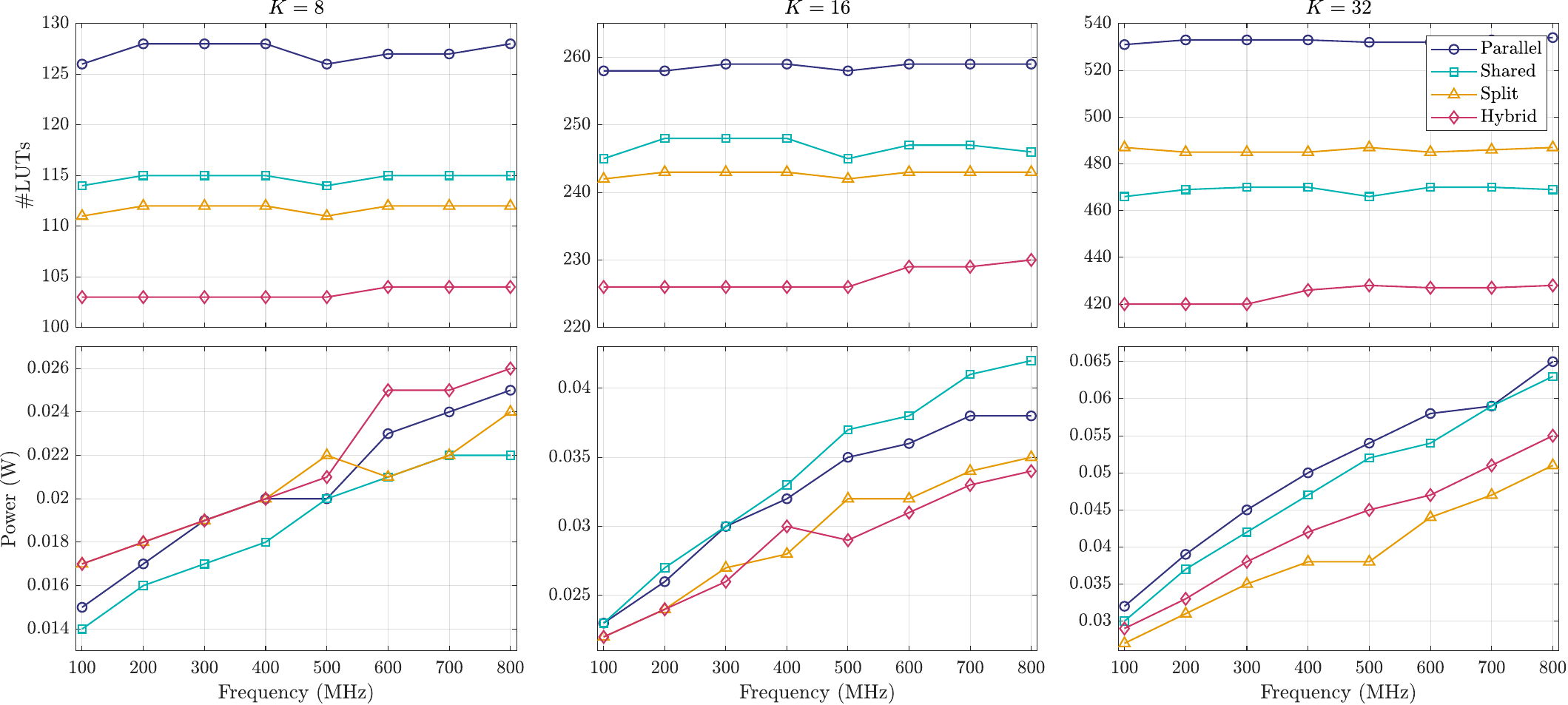}
    \caption{LUT usage and power consumption versus clock frequency for different proposed techniques  with $K = 8, 16,$ and $32$. }
    \label{fig7}
\end{figure*}

\begin{figure}
    \centering    
    \includegraphics[width=0.68\linewidth]{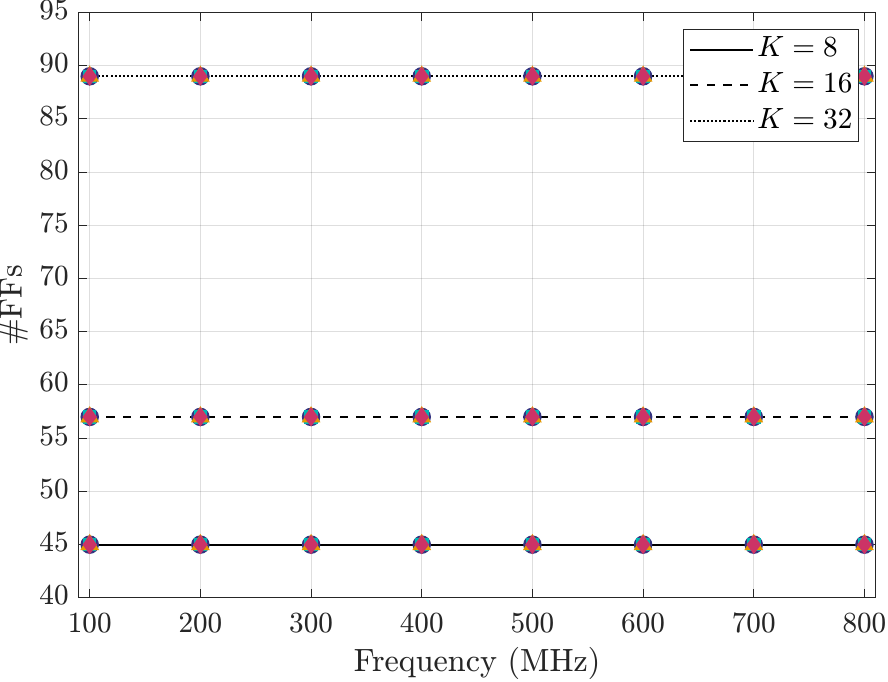}
    \caption{FF usage versus clock frequency for different techniques for $K = 8, 16,$ and $32$. The same markers are used as in Fig. 7.}
    \label{fig8}
\end{figure}

\subsubsection{LUT and FF Usage vs Clock Frequency}
All buffers are implemented using Slice RAM instead of BRAM, improving flexibility and making the architecture better suited for edge deployment while maintaining a fully LUT-based design. Fig.~\ref{fig7} shows the resource utilization of the proposed LUT architectures on the Xilinx ZCU106 FPGA across operating frequencies from 100~MHz to 800~MHz in 100~MHz increments. LUT usage remains stable when timing constraints are easily met but rises at higher frequencies due to tool-driven optimizations like logic duplication and register insertion, reflecting synthesis overhead rather than intrinsic design requirements. All designs show LUT usage consistent with Table~\ref{table1}. The Parallel LUT uses the most LUTs, the Shared LUT reduces usage by reusing partial computations, the Split LUT lowers adders but increases logic via twos complementers (slightly affecting $K=32$), and the Hybrid LUT achieves the most efficient usage with optimized adders and minimal multiplexers. Minor variations arise from synthesis heuristics and routing.  

Unlike LUT usage, flip-flops (FF) utilization remains largely consistent across all LUT designs, as shown in Fig.~\ref{fig8}. This is because LUTs are primarily combinational while FFs store feature maps and pipeline data. Increasing $K$ widens the datapath, requiring more FFs for intermediate results; this increase is similar across all architectures since it depends on sequential storage rather than LUT structure.

\subsubsection{Power Consumption vs Clock Frequency}
Accurate power estimation requires realistic activity data, as post-synthesis estimates often miss switching and routing effects in complex LUTs. In Xilinx Vivado, the procedure is: (1) run functional simulation; (2) generate a .saif file with signal toggles; (3) import it into power analysis; (4) combine toggles with utilization, timing, and clock frequency. This captures actual activity and frequency behavior. Power increases with clock frequency for all LUTs due to higher switching. The slope varies with logic complexity, routing, and active elements: Parallel LUT has the steepest slope from many adders and multiplexers; Shared LUT shows low power at $K=8$ due to computation reuse; Hybrid LUT has slightly higher power at small $K$ but consistent power at $K=16$ and lowest at $K=32$ by pairing adders and minimizing multiplexers; Split LUT shows shallower slopes due to fewer adders. These trends highlight the impact of logic complexity and switching/data reuse on power.

\section{Evaluation of the OBC-GEMM Module}

\begin{figure*}
	\centering    
	\includegraphics[width=0.94\linewidth]{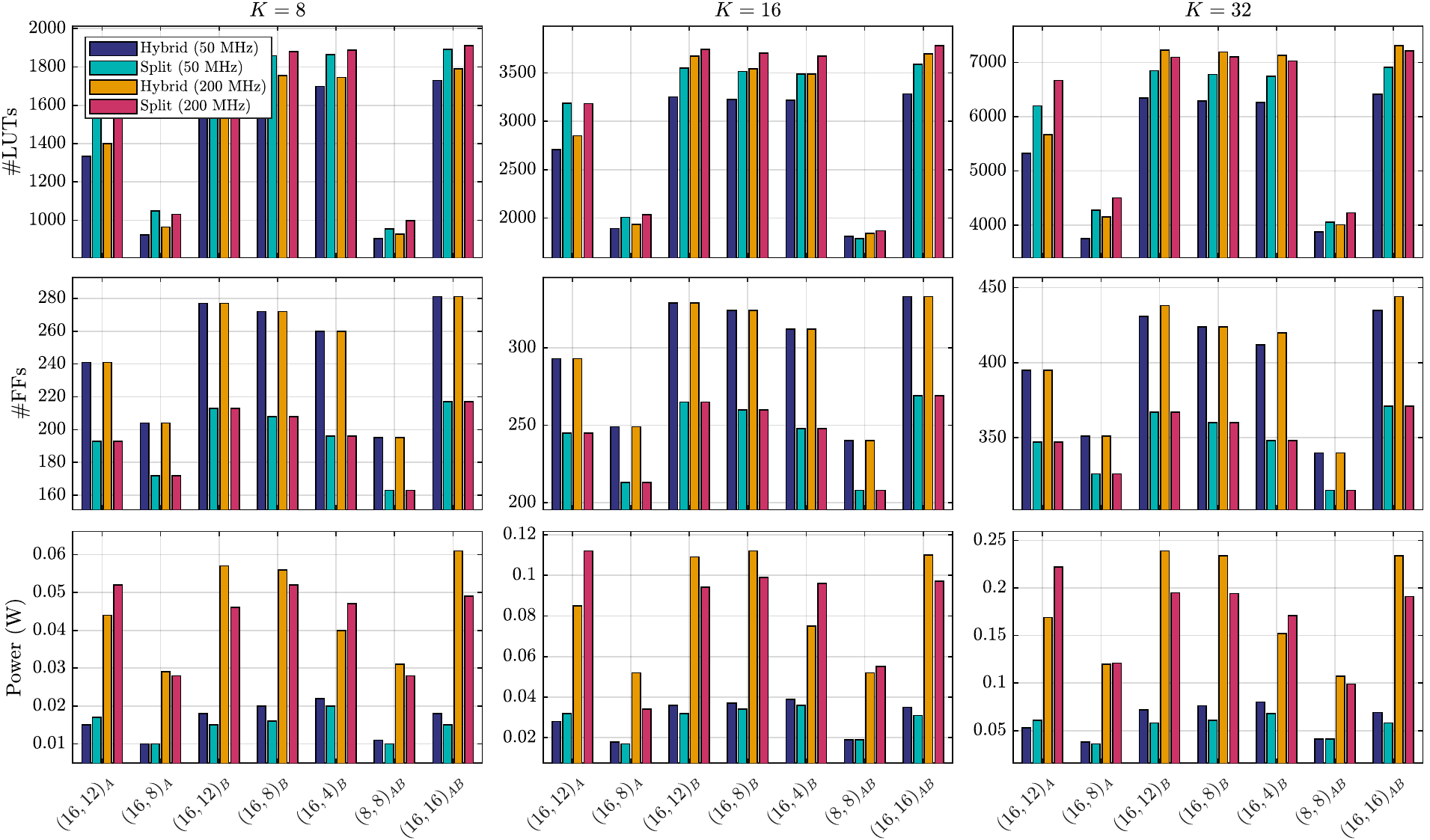}
	\caption{LUTs/FFs usage and power consumption of OBC-GEMM for different $(B_1, B_2)_S$ with $S \in \{A, B, AB\}$ indicates Scheme A or B or Symmetric case - AB values with hybrid and split LUT techniques at 50 MHz and 200 MHz.}
	\label{fig10}
\end{figure*}

For evaluation, we present the Xilinx ZCU106 results of the OBC-GEMM module with input bitwidth $B_1 = 16$ while varying the weight bitwidth $B_2$; the same analysis can be extended to $B_1 = 8$. Input feature maps, weights, and biases are preloaded into on-chip RAM, and the outputs are written back to the same memory. The proposed OBC-GEMM supports asymmetric quantization under Scheme~A and Scheme~B, as described in Section~II. The selected bitwidth combinations in Fig.~\ref{fig10} are primarily motivated by accuracy considerations, which are analyzed in the next section. In particular, the $(16,4)_{B}$ configuration is retained to illustrate the behavior of Scheme~B under aggressive weight quantization, whereas the $(16,4)$ configuration under Scheme~A is omitted due to severe accuracy degradation.

Under asymmetric quantization, both proposed schemes follow $B_1 > B_2$. In Scheme~A, the SA unit operates for $B_1$ clock cycles corresponding to the input bits, while the partial products of the weights are generated using efficient hardware techniques, reducing arithmetic complexity and overall hardware resource usage. In contrast, Scheme~B (Fig.~\ref{fig3c}) bit-slices the weights with bit-width $B_2$ through a PISO for address generation, while inputs in OBC form are used to generate the LUT data. Here, the data remain in the SA unit for $B_2$ clock cycles, enabling faster output, but the wider LUT data ($B_1$ bits) increase logic utilization and power consumption. The $(16,4)_{B}$ case illustrates this trade-off in its most extreme form: very short SA operation but significantly wider LUT paths. 

After integrating the Split and Hybrid LUTs and implementing both schemes, the results are summarized in Fig.~\ref{fig10}. To evaluate the effect of timing constraints, we vary $B_2$, and $K$ at two operating frequencies: 50~MHz and 200~MHz. At 50~MHz, the designs meet timing with minimal replication, while at 200~MHz additional logic and routing are required to satisfy stricter constraints. Under both conditions, Scheme~A and Scheme~B exhibit consistent resource--performance trade-offs, and the choice of $(B_1,B_2)$ directly governs efficiency. The detailed effects on LUT usage, FF count, and power consumption are discussed below.  

\subsubsection{LUT Usage vs. Clock Frequency}
Fig.~\ref{fig10} reports the LUT utilization of the OBC-GEMM module with Split and Hybrid LUTs under different $(B_1,B_2)$ bitwidth pairs at 50~MHz and 200~MHz. Symmetric quantization occurs when $B_1 = B_2$, i.e., $(8,8)_{AB}$ and $(16,16)_{AB}$, where the subscript $AB$ denotes the symmetric case with equal bitwidths for inputs and weights. Among these, $(16,16)_{AB}$ consumes the most LUTs, reflecting the high cost of fully 16-bit datapaths. Asymmetric input-dominant cases, such as $(16,12)_{A}$ and $(16,8)_{A}$, also exhibit substantial LUT demand. However, their usage remains lower than the symmetric baseline since the LUT width is limited by $B_2$. In contrast, Scheme~B configurations, such as $(16,12)_{B}$, $(16,8)_{B}$, and $(16,4)_{B}$, 
incur higher LUT usage than their input counterparts, underscoring the stronger influence of weight precision in determining LUT cost compared to Scheme~A. In fact, $(16,4)_{B}$ shows that pushing $B_2$ down to 4 substantially increases LUT pressure, since the LUT must still encode 16-bit input slices. 
Across all configurations, the Hybrid LUT design consistently requires fewer LUTs than the Split design. Hybrid packing achieves more efficient implementation by sharing PPs, while Split designs incur additional overhead from duplicated logic and reduced computation sharing. This efficiency gap becomes particularly pronounced at higher precision.  

\subsubsection{FF Usage vs. Clock Frequency}
FF utilization, shown in Fig.~\ref{fig10}, demonstrates a different trend compared to LUTs. Unlike LUT usage, FF demand is largely insensitive to clock frequency, remaining relatively stable between 50~MHz and 200~MHz. Instead, FF usage is primarily dictated by datapath width. Asymmetric configurations follow the expected scaling: pairs involving a 16-bit operand, such as $(16,12)_{A}$, $(16,8)_{A}$, $(16,12)_{B}$, $(16,8)_{B}$, and $(16,4)_{B}$, 
consistently fall between the symmetric and fully narrow cases. Notably, $(16,4)_{B}$ requires more FFs than its Scheme~A counterparts of similar width, again reflecting the heavier storage overhead of LUT-based input expansion. 
Architecturally, the Split design consumes more FFs across all precision pairs, owing to the need for wider intermediate registers in its accumulation stage. In contrast, Hybrid LUT avoids unnecessary duplication and demonstrates greater efficiency, especially for mid-range precisions such as $(12,8)_{A}$, $(12,8)_{B}$ and $(8,8)_{AB}$.  

\begin{table*}[t]
    \caption{Description of layer types, channel counts, and activation functions in LeNet-5 {and All-CNN-C}}\label{tab:lenet_layer_mod}
    \centering
    \resizebox{0.90\linewidth}{!}{%
        \begin{threeparttable}
            \begin{tabular}{l|l|l|l|l}
                \hline
                \textbf{Layer} & \multicolumn{2}{c|}{\textbf{LeNet-5 \cite{lecun1998gradient}}} & \textbf{Layer} & \multicolumn{1}{c}{\textbf{All-CNN-C \cite{springenberg2014striving}}}  \\ \cline{2-3} 
                               & \textbf{Original} & \textbf{Modified} & &  \\ \hline
                Conv1           & $5\times5$ conv, 6 ch, tanh         & $5\times5$ conv, 6 ch, ReLU         & {Conv1}  & {$3\times3$ conv, 96 ch, ReLU} \\ \hline 
                Pool1           & $2\times2$ avg pool, stride 2, tanh & $3\times3$ conv, stride 2, ReLU     & {Conv2}  & {$3\times3$ conv, 96 ch, ReLU} \\ \hline
                Conv2           & $5\times5$ conv, 16 ch, tanh        & $5\times5$ conv, 16 ch, ReLU        & {Down1}  & {$3\times3$ conv, 96 ch, stride 2, ReLU} \\ \hline
                Pool2           & $2\times2$ avg pool, stride 2, tanh & $3\times3$ conv, stride 2, ReLU     & {Conv3}  & {$3\times3$ conv, 192 ch, ReLU} \\ \hline
                GAP             & ---                                 & global avg pool on 16 ch            & {Conv4}  & {$3\times3$ conv, 192 ch, ReLU} \\ \hline
                FC1             & dense $400\!\to\!120$, tanh         & dense $16\!\to\!32$, ReLU           & {Down2}  & {$3\times3$ conv, 192 ch, stride 2, ReLU} \\ \hline
                FC2             & dense $120\!\to\!84$, tanh          & dense $32\!\to\!10$, softmax        & {Conv5}  & {$3\times3$ conv, 192 ch, ReLU; $1\times1$ conv, 192 ch, ReLU} \\ \hline
                FC3             & dense $84\!\to\!10$, softmax        & ---                                 & {Global} & {$1\times1$ conv, 10 ch, ReLU; global avg pool; softmax} \\ \hline
            \end{tabular}
            \begin{tablenotes}
                \item Channel counts preserved where applicable; FC layer dimensions are reduced to match modified spatial dimensions.
                \item {The All-CNN-C column summarizes the standard CIFAR-style architecture by stage, where downsampling is performed using stride-2 convolution instead of pooling.}
            \end{tablenotes}
        \end{threeparttable}
    }
\end{table*}
\subsubsection{Power vs. Clock Frequency}
Power consumption, summarized in Fig.~\ref{fig10}, increases noticeably with clock frequency across all configurations. The higher activity and deeper pipelining required to meet 200~MHz timing lead to significantly greater switching energy compared to 50~MHz. Furthermore, the precision scaling strongly impacts power consumption: the widest datapaths, including $(16,16)_{AB}$, $(16,12)_{A}$, and $(16,8)_{A}$, consistently incur the highest power consumption, while cases such as $(8,8)_{AB}$ are far more energy-efficient. Asymmetric configurations reveal an important tradeoff: e.g., $(16,8)_{A}$ requires more clock cycles but lower bits for generating the LUT contents in Scheme~A, resulting in lower power compared to $(16,8)_{B}$, whereas in Scheme~B, dominate switching activity despite shorter bit-width. The $(16,4)_{B}$ case amplifies this effect: the short cycle count reduces SA activity, but the extremely wide LUT data path drives high dynamic power, especially in the Split implementation. 
Between the two architectures, the Hybrid LUT design consistently consumes less power than the Split design. This difference is modest for shorter precisions but becomes substantial for wider datapaths. At $(16,16)_{AB}$ under 200~MHz operation, the Split architecture exhibits a steep increase in power due to redundant toggling in duplicated logic paths, while the Hybrid architecture maintains more controlled energy scaling.  

Symmetric 16-bit quantization $(16,16)_{AB}$ is the most resource- and power-intensive, while narrower datapaths such as $(8,8)_{AB}$ improve energy and area efficiency. Asymmetric cases offer trade-offs: in Scheme~A, logic and power consumption are reduced at the cost of requiring more cycles, whereas in Scheme~B the situation is reversed. The inclusion of $(16,4)_{B}$ demonstrates the extreme of this trend---minimal cycles but substantially higher LUT and power cost---emphasizing the design space boundaries for OBC-GEMM. 
In all cases, the Hybrid architecture outperforms the Split architecture, saving LUTs, FFs, and power---especially at higher precision---thereby demonstrating the scalability of OBC-GEMM across bit-widths.

\section{Models Consideration and Accuracy Results}
\subsection{Models Consideration}
{To comprehensively evaluate the applicability of the proposed OBC-GEMM framework, we consider both a classical LeNet-5} \cite{lecun1998gradient} {and a modern All-CNN-C} \cite{springenberg2014striving} {thereby demonstrating effectiveness across shallow and deeper convolution-dominant workloads. However, the classical LeNet-5 does require lightweight architectural refinements} \cite{imen2022fast}, as summarized in Table~\ref{tab:lenet_layer_mod}. These modifications are primarily motivated by hardware regularity, arithmetic intensity, and fixed-point robustness. Specifically:

\begin{itemize}
    \item Replace \emph{tanh} with ReLU to avoid saturation and simplify quantized implementation.
    \item Replace $2\times2$ average pooling with $3\times3$ stride-2 convolutions to unify computation and improve data reuse.
    \item Replace the two large fully connected (FC) layers with Global Average Pooling (GAP) and a 32-unit FC layer to reduce parameters and memory traffic.
    \item Insert one-side zero padding before each strided convolution to maintain spatial consistency and simplify buffering.
\end{itemize}

These modifications have minimal impact on throughput. Stride-2 convolutions increase total MACs per inference but do not lower achievable MACs per cycle. ReLU and zero-padding add negligible compute cost, while GAP reduces parameters and memory traffic without altering the core computational structure. Thus, the main benefits are improved accuracy robustness and memory efficiency rather than increased compute throughput.

{We further examine All-CNN-C architecture that aligns naturally with the proposed OBC-GEMM core as listed in} Table~\ref{tab:lenet_layer_mod}. {This architecture eliminates pooling and dense layers entirely, relying on repeated $3\times3$ convolutions and stride-2 downsampling. From a hardware perspective, its key characteristics can be summarized as follows}:
\begin{itemize}
    \item {Downsampling is performed exclusively using stride-2 convolutions, ensuring that all layers map uniformly to GEMM operations.}
    \item {Small $3\times3$ kernels are used throughout, increasing nonlinearity depth while maintaining regular memory access patterns.}
    \item {FC layers are replaced with $1\times1$ convolutions followed by GAP and softmax, significantly reducing parameter storage.}
    \item {ReLU is used consistently, improving fixed-point compatibility and eliminating saturation effects.}
    \item {The repetitive convolutional structure simplifies tiling, buffering, and control logic on FPGA.}
\end{itemize}

{As with the modified LeNet-5, these architectural choices primarily influence arithmetic intensity and memory behavior rather than peak MAC/cycle capability. By removing heterogeneous operators such as pooling and large FC layers, All-CNN-C provides a structurally uniform workload that cleanly exercises the convolution-centric datapath of the proposed design.}

\subsection{Accuracy Comparison Under Quantization}
\begin{figure}
	\centering    
	\includegraphics[width=0.81\linewidth]{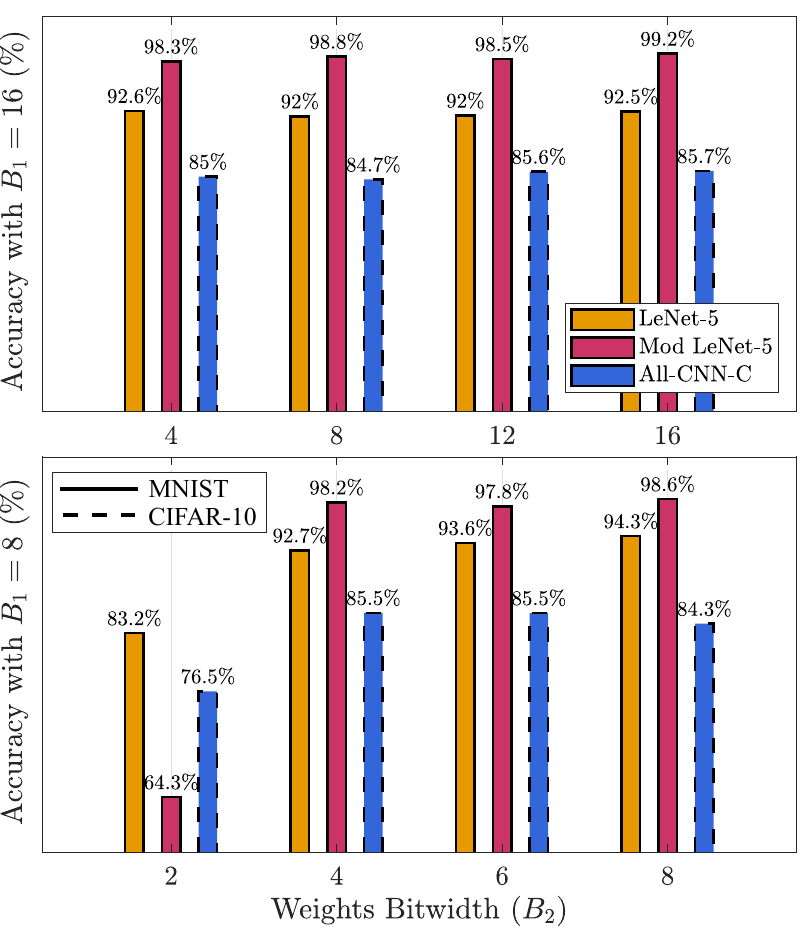}
	\caption{Accuracy of original/modified LeNet-5 {and All-CNN-C} under QAT with two sets of input bit-widths ($B_1$) and varying weight bit-widths ($B_2$).}
	\label{fig9}
\end{figure}

Prior to the FPGA implementation, we first assessed the impact of quantization-aware training (QAT) by comparing the original and modified LeNet-5 on MNIST with padded $32\times32$ inputs.
Both models were trained using fixed-point arithmetic, with input bitwidth $B_1 \in \{16,8\}$ and varying weight bitwidth $B_2$, as shown in Fig.~\ref{fig9}. When the input precision is high ($B_1=16$), the original LeNet-5 achieves $91.96\%$--$92.58\%$ accuracy and shows minimal sensitivity to $B_2$. In contrast, the modified architecture consistently outperforms the baseline across all weight precisions, reaching $98.28\%$ at $B_2=4$ and $99.16\%$ at $B_2=16$. The largest gains occur at moderate weight precisions, indicating that ReLU activations, convolution-based downsampling, and GAP significantly improve robustness to weight quantization and enable near floating-point accuracy even at reduced bitwidth. Under reduced input precision ($B_1=8$), quantization effects become more pronounced. The original network drops to $83.18\%$ at $B_2=2$ but gradually recovers to $94.34\%$ at $B_2=8$. The modified model exhibits greater degradation at extreme low precision e.g., $64.32\%$ at $B_2=2$, yet performance rebounds rapidly, reaching $98.22\%$ at 4 bits and $98.60\%$ at 8 bits. At moderate precision, the modified architecture maintains a clear accuracy advantage, demonstrating strong representational capacity once a modest precision threshold is met.

{We next evaluated All-CNN-C under the same QAT framework on CIFAR-10, using RGB $32 \times 32$ inputs. For $B_1 = 16$, the network achieves nearly identical accuracy for weight bitwidths $B_2 = 4, 8, 12,$ and $16$, indicating limited sensitivity to weight precision when activation bitwidth is high. Increasing $B_2$ beyond 4 bits yields only marginal improvements. When $B_1 = 8$, performance drops to 76.50\% at $B_2 = 2$, but recovers sharply to 85.51\% at 4 bits and remains close to 85\% for $B_2 = 6$--8 bits. This trend is consistent with the LeNet-5 results: ultra-low precision significantly degrades performance, whereas moderate precision is sufficient to stabilize both training and inference. Overall, these results confirm that All-CNN-C maintains stable behavior at $B_2 \ge 4$ bits on CIFAR-10, supporting its suitability for the proposed convolution-centric OBC-GEMM framework. Across both datasets—MNIST for LeNet-5 and CIFAR-10 for All-CNN-C—moderate precision (4–8 bits) offers an effective balance between accuracy and hardware efficiency.}

\section{Performance Evaluation and Comparison}
\subsection{Performance Metric Definitions}
The performance of ML accelerators is commonly expressed in terms of throughput, measured as operations per second (OP/s). Specifically, $\mathcal{T}_{\textnormal{MAC}}$ denotes the number of MAC operations per second. Since a MAC corresponds to a fused multiply–add, this convention is widely adopted for CNN workloads. From (3), throughput is expressed as
\begin{equation}\label{eq13}
	\mathcal{T}_{\textnormal{MAC}} = {\# \text{MACs}}/{T_{s}} = {CKL}/{T_{s}}, 
\end{equation}
where $T_{s}$ is the sample period, assuming all MACs are available on hardware. Under the \textit{im2col} transformation, these MACs can be equivalently expressed as $CL$ IPCs of size $K$:
\begin{equation}\label{eq13a}
	\mathcal{T}_{\textnormal{IPC}} = {\# \text{IPC}}/{T_{s}} = {CL}/{T_{s}}. 
\end{equation}
In our work, all CNN computations are mapped onto the GEMM core; thus, CNN throughput is equivalent to GEMM throughput, expressed in IPCs as in (\ref{eq13a}). The proposed GEMM core instantiates only $L$ IPC blocks based on the OBC-DA (shown in Fig.~\ref{fig3c}), leading to the effective throughput
\begin{equation}\label{eq13b}
	\tilde{\mathcal{T}}_{\textnormal{IPC}} = {\# {\text{IPC}}}/{CT_{s}} = {L}/{T_{s}}. 
\end{equation}
According to the OBC-DA principle, each IPC requires $B$ clock cycles for computation, independent of $K$~\cite{yalamarthy2019low}. This is because partial products are generated while the SA unit produces outputs within $B$ cycles. Consequently, the sample rate $f_s$ must be scaled by $B$ (either $B_1$ for Scheme A or $B_2$ for Scheme B) relative to the clock frequency $f_{clk}$ to maintain synchronization in bit-serial processing. Using $T_s=1/f_s$ with $f_s=f_{clk}/B$, (\ref{eq13b}) simplifies to
\begin{equation}\label{eq15}
	\tilde{\mathcal{T}}_{\textnormal{IPC}} = Lf_s = (L/B)f_{clk}. 
\end{equation}
For fair comparison with MAC-based implementations, $\tilde{\mathcal{T}}_{\textnormal{IPC}}$ is translated to its MAC equivalent by multiplying with $K$, yielding
\begin{equation}\label{eq13c}
	\tilde{\mathcal{T}}_{\textnormal{MAC}} = ({KL}/{B})f_{clk}. 
\end{equation}

The equivalent number of slices (ENS) metric~\cite{liu2019optimized} provides a standardized measure to compare heterogeneous FPGA resources by converting them into a single, unified value. This allows fair and consistent evaluation across different FPGA families and designs. The ENS is defined as
\begin{equation}
	\textnormal{ENS} = {\text{LUT}}/{4} + \text{DSP} \times 102.4 + \text{BRAM} \times 116.2,
\end{equation}
where LUTs are scaled to slices, and DSPs and BRAMs are weighted according to their approximate logic-equivalent cost. Typically, for Xilinx devices, a single DSP block (25$\times$18 multiplier) is equivalent to 128 slices; to account for partial utilization, a factor of 0.8 is applied, yielding 102.4 slices per DSP. Similarly, an 18k BRAM is approximated as 116.2 slices (166 $\times$ 0.7), and an 8k BRAM as 56 slices (70 $\times$ 0.8), reflecting typical dual-ported RAM usage. By converting DSPs and BRAMs into slice-equivalents, ENS provides a unified abstraction that allows designers to quantify resource usage independent of the underlying FPGA family. This simplification highlights trade-offs between logic, computation, and memory, making it suitable for fair cross-platform comparisons of FPGA implementations.

The energy per sample (EPS) is an efficiency metric that captures the trade-off between power consumption and computational throughput. It is defined as
\begin{equation}
	\textnormal{EPS} = {\text{Power}}/{\mathcal{T}_{\textnormal{MAC}}},
\end{equation}
By normalizing power against throughput, EPS quantifies the average energy required to process a single data sample. This measure enables fair comparison of different hardware implementations, as it reflects both computational performance and energy efficiency, which are critical for resource-constrained and high-performance FPGA applications.

Finally, to isolate architectural efficiency from clock frequency scaling, 
we normalize the throughput by both ENS and clock rate, as per 
\begin{equation}\label{eq18}
	\textnormal{AEP} = {\mathcal{T}_{\textnormal{MAC}}}/({f_{clk} \times \textnormal{ENS}})
\end{equation}
where AEP denotes the Architectural Efficiency per ENS. This metric highlights the intrinsic efficiency of the architecture rather than gains achieved simply from higher clock frequency. By using ENS instead of DSP count, AEP captures the contribution of heterogeneous FPGA resources in a unified manner, ensuring that designs which rely heavily on BRAMs or LUTs are fairly compared against DSP-centric implementations. This normalization is particularly important when surveying prior work, where resource usage may vary widely, as it provides a balanced measure of architectural efficiency across different platforms and mapping strategies.

\begin{table*}[t]
\caption{Performance Comparison of Original/Modified LeNet-5 {and All-CNN-C Models} with State-of-the-Art CNNs on FPGA}
\label{tab:perf_cmp}
\centering
\resizebox{1\linewidth}{!}{%
\begin{threeparttable}
\begin{tabular}{l|l|c|c|c|c|c|c|c|c|c|c|c|c}
\hline
\parbox[t]{3mm}{\multirow{19}{*}{\rotatebox[origin=c]{90}{LeNet-5}}} & {\textbf{Design}} & \textbf{BW} & \textbf{Platform} & {\textbf{Freq}} (MHz) & \textbf{LUTs} & \textbf{FFs} & \textbf{DSPs} & \textbf{BRAM} & {\textbf{Power}} (W) & $\mathbf{\mathcal{T}_{\textnormal{MAC}}}$ (GOP/s) & \textbf{ENS} & \textbf{EPS} & \textbf{AEP} \\ \cline{1-14}
&TCS-II~\cite{10472635} & 12/12 & ZYBO Z7 & 100 & 17420 & 10195 & 72 & 80 & -- & 19.20 & 21024.00 & -- & 9.13 \\ \cline{2-14}
&Access~\cite{youssef2022energy} & 8/8 & ZCU106 & 150 & -- & -- & -- & -- & 21.2 & 1.04 & -- & 20.33 & -- \\ \cline{2-14}
&TVLSI~\cite{9923417} & 8/8 & XCVU9P & 300 & 514000 & -- & 512 & 1024 & -- & 1490.50 & 299918.80 & -- & 16.57 \\ \cline{2-14}
&Elect~\cite{cho2021fpga} & 12/8 & Zynq-7020 & 100 & 57657 & 28311 & 123 & 102 & 1.598 & 0.14 & 38861.85 & 11.33 & 0.04 \\ \cline{2-14}
&A. Sci.~\cite{li2023research} & - & Zynq-MPSoC & 100 & 18421 & 17472 & 97 & 13.5 & 3.22 & 0.08 & 16106.85 & 40.76 & 0.05 \\ \cline{2-14}
&Elect.~\cite{ji2023fpqnet} & 8/8 & Alpha Data 9H7 & 250 & 53292 & -- & 2614 & 341 & -- & 110.80 & 320469.80 & -- & 1.38 \\ \cline{2-14}
&{Hybrid$_{\textnormal{AB}}$}  & {8/8} & \multirow{12}{*}{{ZCU106}} & {90}  & {53271}  & {613} & {0} & {0} & {1.44} & {0.18} & {13317.75} & {8.00}  & {0.15} \\ \cline{2-3} \cline{5-14}
&{Hybrid$_{\textnormal{A}}$} & {8/4} &  & {90} & {46722} & {592} & {0} & {0} & {1.43} & {0.18} & {11680.50} & {7.94} & {0.17} \\ \cline{2-3} \cline{5-14}
&{Hybrid$_{\textnormal{B}}$} & {8/4} &  & {90}  & {65682} & {592} & {0}  & {0} & {1.50} & {0.36} & {16420.50} & {4.17}  & {0.24}  \\ \cline{2-3} \cline{5-14}
& {Split$_{\textnormal{AB}}$} & {8/8} &  & {90}  & {52351} & {609} & {0} & {0}  & {1.432} & {0.18} & {13087.75} & {7.96} & {0.15} \\ \cline{2-3} \cline{5-14}
&{Split$_{\textnormal{A}}$} & {8/4} &  & {90} & {46588} & {580} & {0}  & {0}  & {1.41} & {0.18} & {11647.00} & {7.83}  & {0.17}  \\ \cline{2-3} \cline{5-14}
&{Split$_{\textnormal{B}}$} & {8/4} &  & {90}  & {65671} & {580} & {0} & {0} & {1.52} & {0.36} & {16417.75}  & {4.22} & {0.24}  \\ \cline{2-3} \cline{5-14}
&$^{\clubsuit}$Hybrid$_{\textnormal{AB}}$  & 8/8 &  & 100 & 16406 & 1177 & 0 & 0 & 0.835 & 0.20 & 4101.50 & 4.18 & 0.49 \\ \cline{2-3} \cline{5-14}
&$^{\clubsuit}$Hybrid$_{\textnormal{A}}$ & 8/4 &  & 100 & 14724 & 1044 & 0 & 0 & 0.824 & 0.20 & 3681.00 & 4.12 & 0.54 \\ \cline{2-3} \cline{5-14}
&$^{\clubsuit}$Hybrid$_{\textnormal{B}}$ & 8/4 &  & 95 & 23019 & 1043 & 0 & 0 & 0.976 & 0.38 & 5754.75 & 2.57 & 0.70 \\ \cline{2-3} \cline{5-14}
&$^{\clubsuit}$Split$_{\textnormal{AB}}$ & 8/8 &  & 100 & 16310 & 1170 & 0 & 0 & 0.832 & 0.20 & 4077.50 & 4.16 & 0.49 \\ \cline{2-3} \cline{5-14}
&$^{\clubsuit}$Split$_{\textnormal{A}}$& 8/4 &  & 100 & 14741 & 1041 & 0 & 0 & 0.826 & 0.20 & 3685.25 & 4.13 & 0.54 \\ \cline{2-3} \cline{5-14}
&$^{\clubsuit}$Split$_{\textnormal{B}}$ & 8/4 &  & 95 & 23071 & 1041 & 0 & 0 & 0.949 & 0.38 & 5767.75 & 2.50 & 0.69 \\ \hline

\parbox[t]{3mm}{\multirow{15}{*}{\rotatebox[origin=c]{90}{All-CNN-C}}}
& {CDASI} \cite{meloni2019cnn} & {16/8} & {Zynq-XC7Z007S} & {80} & {12610}  & {12501} & {54} & {44} & {$\sim$ 2} & {10.62} & {13794.90} & {0.19} & {9.62}\\ \cline{2-14}

& MAC Baseline & {8/8} & \multirow{8}{*}{{XCZU21DR}}
& {95} 
& {59648} 
& {1228} 
& {27} 
& {64} 
& {1.662} 
& {0.19} 
& {25113.60} 
& {8.75} 
& {0.08} 
\\ \cline{2-3} \cline{5-14}

& MAC Baseline & {8/4} & 
& {95} 
& {59086} 
& {1081} 
& {27} 
& {32} 
& {1.598} 
& {0.38} 
& {21254.70} 
& {4.21} 
& {0.19} \\ \cline{2-3} \cline{5-14}

&{Hybrid$_{\textnormal{AB}}$}  & {8/8} & 
& {50} 
& {328033} 
& {870} 
& {0} 
& {0} 
& {4.35} 
& {0.10} 
& {82008.25} 
& {43.50} 
& {0.02} 
\\ \cline{2-3} \cline{5-14}

&{Hybrid$_{\textnormal{A}}$} & {8/8} & 
& {50} 
& {253813} 
& {738} 
& {0} 
& {0} 
& {2.97} 
& {0.10} 
& {63453.25} 
& {29.70} 
& {0.03} 
\\ \cline{2-3} \cline{5-14}

& {Hybrid$_{\textnormal{B}}$} & {8/4} & 
& {50} 
& {384078} 
& {617} 
& {0} 
& {0} 
& {4.98} 
& {0.19} 
& {96019.50} 
& {26.21} 
& {0.04} 
\\ \cline{2-3} \cline{5-14}

& {Split$_{\textnormal{AB}}$} & {8/8} & 
& {50} 
& {199367} 
& {742} 
& {0} 
& {0} 
& {3.23} 
& {0.08} 
& {49841.75} 
& {43.07} 
& {0.03} 
\\ \cline{2-3} \cline{5-14}

& {Split$_{\textnormal{A}}$} & {8/4} & 
& {50} 
& {253650} 
& {750} 
& {0} 
& {0} 
& {3.03} 
& {0.10} 
& {63412.50} 
& {30.30} 
& {0.03} 
\\ \cline{2-3} \cline{5-14}

& {Split$_{\textnormal{B}}$} & {8/4} & 
& {50} 
& {384319} 
& {617} 
& {0} 
& {0} 
& {4.90} 
& {0.19} 
& {96079.75} 
& {25.79} 
& {0.04} 
\\  \cline{2-14}

& $^{\spadesuit}${Hybrid$_{\textnormal{AB}}$} & {8/8} & \multirow{6}{*}{{ZCU106}}
& {100}
& {59986}
& {1242}
& {0}
& {64}
& {1.371}
& {0.20}
& {22433.30}
& {6.86}
& {0.09}
\\ \cline{2-3} \cline{5-14}

& $^{\spadesuit}${Hybrid$_{\textnormal{A}}$} & {8/4} & 
& {100}
& {59679}
& {1094}
& {0}
& {32}
& {1.338}
& {0.20}
& {18638.15}
& {6.69}
& {0.11}
\\ \cline{2-3} \cline{5-14}

& $^{\spadesuit}${Hybrid$_{\textnormal{B}}$} & {8/4} & 
& {100}
& {53684}
& {1063}
& {0}
& {64}
& {1.236}
& {0.40}
& {20857.80}
& {3.09}
& {0.19}
\\ \cline{2-3} \cline{5-14}

& $^{\spadesuit}${Split$_{\textnormal{AB}}$} & {8/8} & 
& {100}
& {59971}
& {1241}
& {0}
& {64}
& {1.368}
& {0.20}
& {22429.55}
& {6.84}
& {0.09}
\\ \cline{2-3} \cline{5-14}

& $^{\spadesuit}${Split$_{\textnormal{A}}$} & {8/4} & 
& {100}
& {59705}
& {1094}
& {0}
& {32}
& {1.344}
& {0.20}
& {18644.65}
& {6.72}
& {0.11}
\\ \cline{2-3} \cline{5-14}

& $^{\spadesuit}${Split$_{\textnormal{B}}$} & {8/4} & 
& {100}
& {53844}
& {1063}
& {0}
& {64}
& {1.229}
& {0.40}
& {20897.80}
& {3.07}
& {0.19}
\\ \hline
\end{tabular}
\begin{tablenotes}
\item BW: Bitwidth format (input act./weight), $^{\clubsuit}:$ Modified LeNet-5, $^{\spadesuit}:$ with BRAM, ENS = LUT/4 + DSP $\times 102.4$ + BRAM $\times 116.2$ \cite{liu2019optimized}. Assumptions: 1 DSP = $128\times0.8=102.4$ slices, 1 BRAM (18k) = $166\times0.7=116.2$ slices, 1 slice = 4 LUTs. EPS = Power / $\mathcal{T}_{\textnormal{MAC}}$ (W per GOP/s). AEP = $1000\times \mathcal{T}_{\textnormal{MAC}}/(f_{clk}\times \textnormal{ENS}/1000)$, reported in ops/cycle/kiloENS.
\end{tablenotes}
\end{threeparttable}
}
\end{table*}

\subsection{Evaluation of the Models and their Comparison} 
The proposed OBC-GEMM core is validated on two FPGA platforms: the Xilinx ZCU106 and the XCZU21DR. LeNet-5 is evaluated on the ZCU106, while All-CNN-C is assessed on both platforms to investigate the effect of slice LUT utilisation over BRAM on the deeper network. The hardware-optimized LUT architectures have already been determined through the implementation workflow and are now integrated into the complete CNN models. The resulting accelerator employs Hybrid and Split LUT techniques with Scheme~A and Scheme~B, operating entirely without DSPs or BRAMs. Note that the notations in Table~\ref{tab:perf_cmp} are defined as follows: Hybrid$_{\textnormal{A}}$/Split$_{\textnormal{A}}$ denote asymmetric quantization with Scheme~A, Hybrid$_{\textnormal{B}}$/Split$_{\textnormal{B}}$ denote asymmetric quantization with Scheme~B, and Hybrid$_{\textnormal{AB}}$/Split$_{\textnormal{AB}}$ denote symmetric quantization applied across all layers.

The original LeNet-5 maintains a consistent 90~MHz across all variants. For the modified LeNet-5 on the ZCU106, Scheme~A variants operate at 100~MHz, while Scheme~B variants operate at 95~MHz.  Both the original and modified LeNet-5 designs eliminate DSPs and BRAMs, relying solely on slice LUTs.\ The modified LeNet-5 achieves a substantial reduction in LUT utilisation and power consumption (0.824--0.976~W vs.\ 1.41--1.52~W), though at the cost of a modest increase in FF count. Throughput remains comparable across both variants, while Scheme~B provides higher computational speed than Scheme~A, at the cost of higher LUT usage and greater bandwidth demand to sustain the increased processing rate.
The modified LeNet-5 demonstrates improved energy efficiency with lower EPS values (Hybrid and Split: 2.57--4.18 vs.\ 4.17--8.00~W/GOP/s), while it also achieves higher AEP (Hybrid and Split: 0.49--0.70 vs.\ 0.15--0.24~ops/cycle/kENS), indicating better throughput per cycle relative to its resource footprint. It is therefore evident that the modified LeNet-5 offers a more favourable balance between resource utilisation and energy efficiency, making it better suited for resource-constrained deployment when implemented using slice LUTs only. Table~\ref{tab:perf_cmp}  also compares the original and modified LeNet-5 designs with state-of-the-art implementations. Prior work such as~\cite{9923417} achieves very high throughput of 1490.5~GOP/s at 300~MHz by relying heavily on DSPs and BRAMs, while smaller-scale implementations such as~\cite{cho2021fpga} provide only modest throughput at considerable resource and energy cost. In contrast, both the original and modified LeNet-5 designs deliver competitive EPS against other MAC-based implementations. While a direct AEP comparison with MAC-based implementations is inherently less straightforward due to the bit-serial nature of OBC schemes, the modified LeNet-5 nevertheless achieves superior AEP over several smaller-scale implementations — demonstrating a favourable balance between computational efficiency and resource cost.

In contrast, All-CNN-C is evaluated under two configurations. For the BRAM-free implementation, the network parameters must be mapped entirely onto slice resources, imposing a significantly higher LUT requirement; therefore, the LUT-richer XCZU21DR is selected, with a maximum frequency of 50~MHz. For the BRAM-enabled implementation, main storage is absorbed by on-chip block memory, significantly reducing LUT pressure; as a result, the design can be implemented on the resource-constrained ZCU106, operating at ~100 MHz. This comparison is used to evaluate whether a slice-LUT-only or BRAM-assisted implementation is more suitable for the deeper All-CNN-C network. The results in Table~\ref{tab:perf_cmp} show that, compared with the slice-LUT-only XCZU21DR configuration, the BRAM-enabled ZCU106 configuration operates at a higher clock frequency with lower power consumption, yielding significantly improved EPS (Hybrid and Split: 3.09--6.86 vs.\ 25.79--43.50~W/GOP/s) and AEP (Hybrid and Split: 0.089--0.192 vs.\ 0.024--0.040~ops/cycle/kENS). This suggests that, as network complexity increases, BRAM utilisation helps alleviate LUT pressure and improve overall energy efficiency~\cite{venieris2016fpgaconvnet}. This contrasts with the LeNet-5 findings, where the slice-LUT-only approach proved sufficient and effective. For All-CNN-C~\cite{springenberg2014striving}, only a limited number of FPGA implementations are available in the literature. The only directly comparable reference is~\cite{meloni2019cnn}, which implements the 
same network but reports an inaccurate power figure of $\sim$2~W from the Xilinx Power Estimator and thus the derived EPS value is indicative rather than definitive.
MAC baselines of All-CNN-C are therefore also realized, employing DSPs and BRAMs, consuming 1.60--1.66~W compared to 1.23--1.37~W for the proposed BRAM-enabled ZCU106 variants. The proposed variants operate entirely without DSPs, yet achieve comparable EPS (Hybrid and Split: 3.09--6.86 vs.\ 4.21--8.75~W/GOP/s) and AEP (0.089--0.192 vs.\ 0.080--0.188~ops/cycle/kENS) to the MAC baselines, demonstrating 
that DSP-free LUT-based inference remains competitive in energy efficiency for deeper networks, despite the inherent routing complexity and clock frequency limitations associated without BRAM exploitation.

\section{{Conclusion}}
We have presented COMET, a co-optimization framework for DSP-free CNN inference on FPGAs using OBC techniques. Inference has been formulated with OBC representations applied separately to inputs (Scheme~A) and weights (Scheme~B), exploiting bit-width asymmetry, while the SA operation incorporates the offset term with pre-scaled bias. OBC-based LUT implementations, including parallel, shared, split, and hybrid, have been evaluated to highlight trade-offs in performance, resource usage, and energy efficiency, and integrated into an OBC-GEMM core via the \textit{im2col} transformation for resource-aware CNN execution. Both the original and modified LeNet-5 have demonstrated the viability of slice-LUT-only CNN inference, with the modified design achieving a more 
favourable balance between resource utilisation and energy efficiency. All-CNN-C has further validated scalability for deeper networks, where BRAM utilisation has been shown to alleviate LUT pressure and improve energy efficiency as network complexity increases. These results collectively demonstrate that DSP-free FPGA deployment is achievable across CNN models of varying complexity, establishing COMET as a scalable and practical solution for LUT-centric CNN inference across edge AI applications.

\bibliographystyle{IEEEtran}
\bibliography{ref}

\begin{IEEEbiography}
[{\includegraphics[width=1.07in,height=1.28in,clip,keepaspectratio]{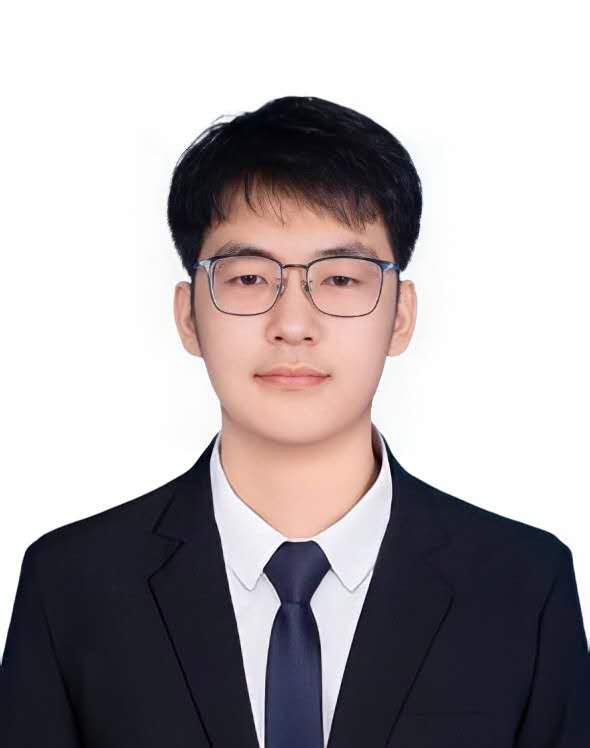}}]
{Boyang Chen} received his B.Eng. degree in Electronics from Heriot-Watt University, UK, and Xidian University, China, in 2025, through a joint undergraduate program. He is currently pursuing an MSc in Communications and Signal Processing at Imperial College London, UK. His research focuses on signal processing and hardware acceleration with FPGA-based architectures, with particular interest in low-level optimization and efficient system-level implementations of signal processing algorithms.
\end{IEEEbiography}
\vfill
\begin{IEEEbiography}
[{\includegraphics[width=1.07in,height=1.28in,clip,keepaspectratio]{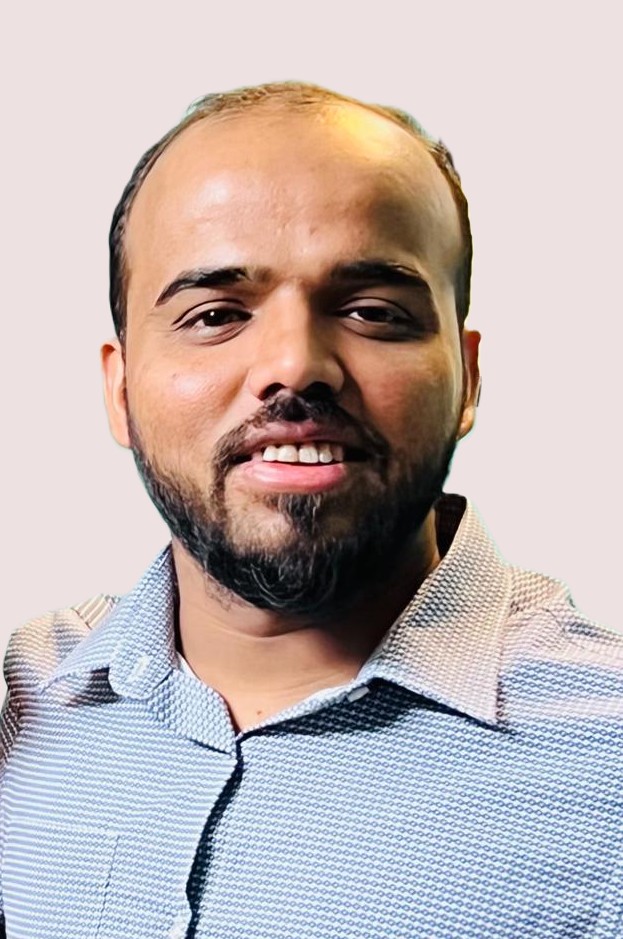}}]
{Mohd Tasleem Khan} (Member, IEEE) received his B.Tech degree in Electronics in 2013 from AMU, India, and his Ph.D. in VLSI in 2019 from IIT Guwahati, India. He was a Principal Engineer at TSMC, Hsinchu, Taiwan, and has worked as an Assistant Professor in the Department of Electronics Engineering at IIT Dhanbad, India. He was a Postdoctoral Research Associate at Linköping University, Sweden, from 2021 to 2024. Currently, he is an Assistant Professor at Heriot-Watt University, Edinburgh, UK. His research and teaching interests include algorithms and architectures for VLSI implementation in Signal Processing, Machine Learning/AI, and Communication Systems. He is currently serving as an Associate Editor for IEEE Signal Processing Letters, IEEE Transactions on Neural Networks and Learning Systems; and IEEE Transactions on Automation Science and Engineering, and is part of the Editorial Board of IEEE Embedded Systems Letters. He is a member of the IEEE Signal Processing Society and the IEEE Circuits and Systems Society. He has served as a TPC member for IEEE ICJNN'25, ISVLSI'25, ICDCS'25.
\end{IEEEbiography}
\vspace{-1.20cm}
\begin{IEEEbiography}
	[{\includegraphics[width=1.07in,height=1.28in,clip,keepaspectratio]{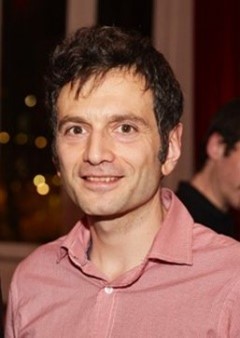}}]
	{George Goussetis} (IEEE Fellow) received the Diploma degree in Electrical and Computer Engineering from the National Technical University of Athens, Greece, in 1998, and the Ph.D. degree from the University of Westminster, London, UK, in 2002. In 2002 he also graduated B.Sc. in physics (first class) from University College London (UCL), UK. In 1998, he joined the Space Engineering, Rome, Italy, as RF Engineer and in 1999 the Wireless Communications Research Group, University of Westminster, UK, as a Research Assistant. Between 2002 and 2006 he was a Senior Research Fellow at Loughborough University, UK. He was Assistant Professor with Heriot-Watt University, Edinburgh, UK between 2006 and 2009 and Associate Professor with Queen’s University Belfast, UK, between 2009 and 2013. In 2013 he joined Heriot-Watt and was promoted to Professor in 2014. He is currently the director of the Institute of Sensors Signals and Systems at Heriot-Watt University. He has authored or co-authored over 500 peer-reviewed papers several book chapters one book and six patents. His research interests are in the area of microwave and antenna components and subsystems. 
	Dr. Goussetis held research fellowships from the Onassis foundation in 2001, the UK Royal Academy of Engineering between 2006-2011, and European Commission Marie-Curie in 2011-12 and again in 2014-17. He is the co-recipient of the 2011 European Space Agency young engineer of the year prize, the 2011 EuCAP best student paper prize, the 2012 EuCAP best antenna theory paper prize and the 2016 Bell Labs prize. He is the co-recipient of the Best Paper Award 2023 for his contributions in IEEE Proceedings. He served as Associate Editor to the IEEE Antennas and Wireless Propagation Letters between 2014-18.
\end{IEEEbiography}
\vfill
\vspace{-1.5cm}
\begin{IEEEbiography}
	[{\includegraphics[width=1.07in,height=1.28in,clip,keepaspectratio]{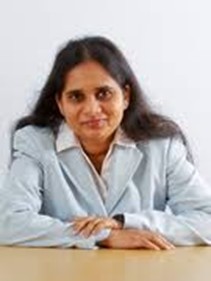}}]
	{Mathini Sellathurai} (IEEE Fellow)  is currently a professor of signal processing and wireless communications and Dean of Science and Engineering at Heriot-Watt University, Edinburgh, U.K. She has been active in signal processing research for the past 20 years and has a strong international track record in wireless communications. She held visiting positions with Bell-Laboratories, Holmdel, NJ, USA, and at the Canadian Communications Research Centre, Ottawa, Canada. She has published over 250 peer-reviewed papers in leading international journals and conferences and two research monographs. Her present research includes machine learning and statistical signal processing, wireless communications, full-duplex communications and radar, assistive care, robotics and hearing aids. She is a recipient of an IEEE Communication Society WIE mentorship award (2022), the Fred W. Ellersick Best Paper Award (2005), the Industry Canada Public Service Awards (2005), and Technology Transfers awards (2004). She received the Natural Sciences and Engineering Research Council of Canada (NSERC) Doctoral Award for her Ph.D. dissertation. She was an Editor for IEEE Transactions of Signal Processing from 2009 to 2018, the General Chair of IEEE Signal Processing Advances in Wireless Communications (2016), and a member of the IEEE SPCOM Technical Committee from 2014 to 2019. She is a member of the IEEE History Committee and Strategic Advisory Team of  UK Research Council.. She is  an invited Fellow of the Asia-Pacific Artificial Intelligence Association (AAIA) and Women Engineering Society (WES), U.K.

\end{IEEEbiography}
\vspace{-1.5cm}

\begin{IEEEbiography}
	[{\includegraphics[width=1.07in,height=1.28in,clip,keepaspectratio]{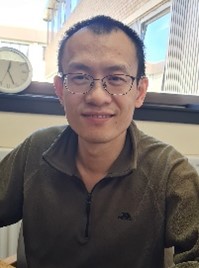}}]
	{Yuan Ding} (Member, IEEE) received the bachelor’s degree in electronic engineering from Beihang University, Beijing, China, in 2004, the master’s degree in electronic engineering from Tsinghua University, Beijing, in 2007, and the Ph.D. degree in electronic engineering from Queen’s University Belfast, Belfast, U.K., in 2014.
	He was a Radio Frequency (RF) Engineer with the Motorola Research and Development Centre, Beijing, from 2007 to 2009, before joining Freescale Semiconductor Inc., Beijing, as an RF Field Application Engineer, responsible for high-power base-station amplifier design, from 2009 to 2011. He is currently an Associate Professor with the Institute of Sensors, Signals and Systems, Heriot-Watt University, Edinburgh, U.K. His research interests are in the IoT-related physical-layer designs, antenna array, physical-layer security, massive active arrays, and other B5G related areas.
	Dr. Ding was the recipient of the IET Best Student Paper Award at LAPC 2013, the Young Scientists Awards in General Assembly and Scientific Symposium at the 2014 XXXIst URSI, Best Paper Award in International Conference on the UK-China Emerging Technologies (UCET), and the co-recipient of the Best Paper Award 2023 for his contributions in IEEE Proceedings.
\end{IEEEbiography}
\vspace{-1.1cm}
\begin{IEEEbiography}[{\includegraphics[width=1in,height=1.25in,clip,keepaspectratio]{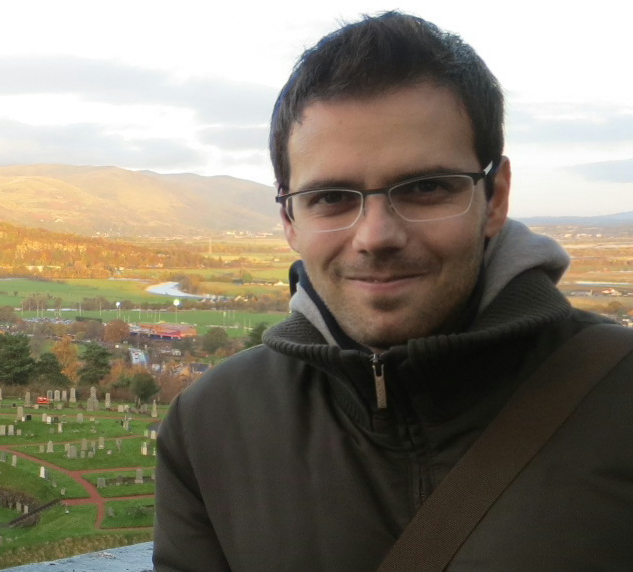}}]{João F. C. Mota}
	received the M.Sc. and Ph.D. degrees in electrical and computer engineering from the Technical University of Lisbon, Lisbon, Portugal, in 2008 and 2013, respectively, and the Ph.D. degree in electrical and computer engineering from Carnegie Mellon University, Pittsburgh, PA, USA, in 2013.
	He is currently an Assistant Professor of Signal and Image Processing with Heriot-Watt University, Edinburgh, U.K.
	His research interests include theoretical and practical aspects of high-dimensional data processing, inverse problems, optimization theory, machine learning, data science, and distributed information processing and control.
	Dr. Mota was a recipient of the 2015 IEEE Signal Processing Society Young Author Best Paper Award and is currently Associate Editor for IEEE Transactions on Signal Processing.
\end{IEEEbiography}
\vspace{-0.8cm}
\begin{IEEEbiography}[{\includegraphics[width=1in,height=1.25in,clip,keepaspectratio]{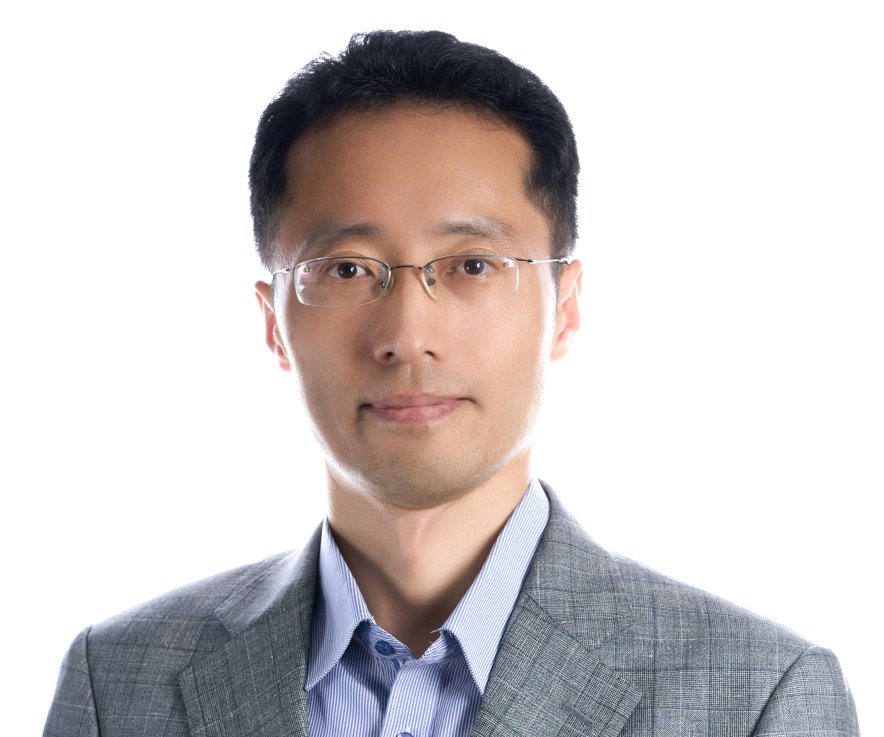}}]{Jongeun Lee}  (S’01-M’11) received the B.S. and
	M.S. degrees in electrical engineering and the Ph.D.
	degree in electrical engineering and computer science from Seoul National University, Seoul, Korea,
	in 1997, 1999, and 2004, respectively.
	Since 2009, he has been with the Faculty of Ulsan 	National Institute of Science and Technology, Ulsan,
	South Korea, where he is a Professor in Electrical 	Engineering. His research interests include neural
	network processors, reconfigurable architectures, in-memory computing, and compilers.
	
\end{IEEEbiography}
\vfill
\ifCLASSOPTIONcompsoc
\else
\fi
\ifCLASSOPTIONcaptionsoff
  \newpage
\fi
\end{document}